\documentclass[aps,prd,preprint,nofootinbib,eqsecnum,superscriptaddress]{revtex4-2}

\usepackage{mathtools}
\usepackage{hyperref}
\begin{document}
\title{Geodesic Focusing Conditions in \texorpdfstring{$f(Q)$}{f(Q)} Gravity}

\author{Chao-Qiang Geng}
\email[Electronic address: ]{cqgeng@ucas.ac.cn}
\affiliation{School of Fundamental Physics and Mathematical Sciences,
Hangzhou Institute for Advanced Study, UCAS, Hangzhou 310024, China}
\affiliation{Synergetic Innovation Center for Quantum Effects and Applications, Hunan Normal University, Changsha 410081, China}
\author{Chunhui Liu}
\email[Electronic address: ]{liuchunhui22@mails.ucas.ac.cn}
\affiliation{School of Fundamental Physics and Mathematical Sciences,
Hangzhou Institute for Advanced Study, UCAS, Hangzhou 310024, China}
\affiliation{University of Chinese Academy of Sciences, Beijing 100190, China}
\affiliation{Institute of Theoretical Physics, Chinese Academy of Sciences, Beijing 100190, China}
\author{Ling-Wei Luo}
\email[Electronic address: ]{lwluo.phys@gmail.com}
\affiliation{School of Fundamental Physics and Mathematical Sciences,
Hangzhou Institute for Advanced Study, UCAS, Hangzhou 310024, China}
\author{Jianhui Qiu}
\email[Electronic address: ]{jhqiu@nao.cas.cn}
\affiliation{School of Fundamental Physics and Mathematical Sciences,
Hangzhou Institute for Advanced Study, UCAS, Hangzhou 310024, China}

\begin{abstract}
We study the geodesic deviation equation in symmetric teleparallel geometry (STG), 
where the relative acceleration is defined with respect to the STG connection. 
We analyze the modified Raychaudhuri equation along a geodesic congruence in $f(Q)$ gravity under the Weyl-type ansatz, together with an additional assumption under which the metric variation term along the congruence is converted into a disformation-induced acceleration term.
In contrast to the purely geometrical Raychaudhuri equation obtained in general metric-affine settings, 
the equation derived here contains matter-source contributions through the trace equation of $f(Q)$ gravity.
Different from general relativity, 
focusing in $f(Q)$ gravity is not automatic, 
and one must impose an appropriate focusing condition. 
We collect the model-dependent terms in the modified Raychaudhuri equation into an effective energy-momentum trace $T_{\text{eff}}$, 
so that the focusing condition can be written as the inequality $T\leq T_{\text{eff}}$, 
where $T$ is the trace of the matter energy-momentum tensor. 
We also apply this condition to the flat Friedmann--Lema\^{i}tre--Robertson--Walker (FLRW) background. 
The homogeneous and isotropic STG connection admits three branches, each characterized by a single connection function $\gamma_i$, with $i=1,2,3$. 
Only the first branch with the coincident gauge is compatible with the Weyl-type ansatz. 
We obtain the resulting effective trace $T_{\text{eff}}=T$ for any form of $f(Q)$ satisfying $f_Q>0$ in the flat FLRW universe. The focusing inequality is saturated and imposes no additional constraint on the matter content.
\end{abstract}


\maketitle
\newpage 

\section{Introduction}
The geometric nature of gravity, as embodied in General Relativity (GR), is one of the most profound paradigms in modern physics. However, this geometric interpretation is not exhausted by curvature alone. 
Historically, Weyl first introduced the nonmetricity in his theory of gravitation and electromagnetism and developed the so-called ``purely infinitesimal'' geometry~\cite{Weyl:1918ib,Weyl:1918pdp}. 
In a general metric-affine framework~\cite{Hehl:1994ue}, the connection should be determined not only by the Levi-Civita connection $\mathring{\Gamma}^{\rho}{}_{\mu\nu}$, but also by torsion $T^{\rho}{}_{\mu\nu}$ and nonmetricity $Q_{\rho\mu\nu}$, and the latter two result in the contortion tensor $K^{\rho}{}_{\mu\nu}$ and the disformation tensor $\mathrm{L}^{\rho}{}_{\mu\nu}$ in the connection. 
Einstein formulated the teleparallelism in a metric theory which developed an equivalence class theory of GR~\cite{Einstein:ap}, in which the torsion tensor is introduced implicitly.
Similarly, another dynamically equivalent formulation of GR constructed by a flat connection with nonmetricity in the metric-affine framework is demonstrated by Nester and Yo~\cite{Nester:1998mp}, 
which is called the symmetric teleparallel equivalent of GR (STEGR) in symmetric teleparallel geometry (STG). 
For the great detail of nonmetricity in the teleparallel geometry, readers can consult recent reviews~\cite{Heisenberg:2018vsk,Bahamonde:2021gfp}.
Recently, the widely studied extension of STEGR is $f(Q)$ gravity~\cite{BeltranJimenez:2017tkd,BeltranJimenez:2018vdo,Heisenberg:2023lru}, 
where the Lagrangian is generalized to a function of the nonmetricity scalar $Q$. 
Although the cosmological applications of $f(Q)$ gravity have been extensively explored~\cite{BeltranJimenez:2019tme,Hohmann:2020zre,Hohmann:2021ast,Heisenberg:2022mbo,Gomes:2023tur,Heisenberg:2023tho}, several kinematic aspects of the theory have also begun to be investigated. 

Geodesic and autoparallel equations are, in general, not the same in non-Riemannian geometry.
The action principles for these two equations in the presence of nonmetricity have recently been demonstrated by Heisenberg~\cite{Heisenberg:2026ess}.
The equation of geodesic deviation in $f(Q)$ gravity and its cosmological applications in the Friedmann--Lema\^{i}tre--Robertson--Walker (FLRW) universe were discussed by Beh, Loo, and De~\cite{Beh:2021wva}. 
The authors directly obtained the geodesic deviation in $f(Q)$ gravity by replacing the tidal term of the equation of geodesic deviation in GR by using the teleparallelism identity of the Riemann curvature. 
In a later work, Capozziello, Capriolo, and Nojiri also derived the deviation equation directly for nearby freely falling particles and made the disformation/nonmetricity contributions explicit~\cite{Capozziello:2024vix}. 
However, their form for the geodesic deviation equation differs from the one given by Beh, Loo, and De~\cite{Beh:2021wva}.

A general treatment of the kinematics of timelike autoparallel congruence with torsion and nonmetricity has been discussed in Refs.~\cite{Iosifidis:2018diy,Agashe:2023vsz}, and the authors derived the modified Raychaudhuri equation at a purely geometric level,
without imposing the gravitational equation of motion. 
Based on this work, Madhukrishna Chakraborty and Subenoy Chakraborty investigated the modified Raychaudhuri equation in homogeneous and isotropic cosmology in $f(Q) = Q + F(Q)$ gravity, and then they studied the possible avoidance of the initial big-bang singularity for polynomial and exponential choices of $F(Q)$~\cite{Chakraborty:2024sco}. 

In fact, there is a subtlety coming from the covariant definition of the acceleration which leads to different physical forms of the geodesic deviation equation.
The shortest path (the geodesic) and the straightest path (the autoparallel), in general, need not coincide in STG. Two paths coincide with each other in Riemannian geometry due to the unique metric connection of Levi-Civita. 
We provide an alternative 
derivation of the geodesic equation and 
its corresponding deviation equation. 
The subsequent analysis of the modified Raychaudhuri equation is based on the kinematics derived from the resulting geodesic deviation equation, together with a Weyl-type ansatz.
We further impose an assumption that the metric variation term along the congruence is converted into a disformation-induced acceleration term.
In our calculations, the modified Raychaudhuri equation is no longer a purely geometric relation, since it is affected by matter distributions.
However, obtaining a focusing theorem in STG is nontrivial.
Instead, we must identify an appropriate focusing condition in our model. 
The focusing condition can be expressed as an inequality involving the trace $T$ of the matter energy-momentum tensor. 
In particular, it can be written as $T\leq T_\text{eff}$, where $T_\text{eff}$ is defined as an effective energy-momentum trace induced by the geometric background. 
Under cosmological symmetry, the flat FLRW background has three branches of the homogeneous and isotropic STG connection in $f(Q)$ gravity~\cite{Hohmann:2020zre, Hohmann:2021ast, Heisenberg:2022mbo}. 
The compatibility between the Weyl-type ansatz and the homogeneous and isotropic STG connection in the flat FLRW universe is studied.
Only the coincident gauge is compatible with the Weyl-type ansatz. 
Under these conditions, we evaluate $T_\text{eff}$ in the flat FLRW background within the coincident gauge and find that $T_\text{eff} = T$ for any function $f(Q)$ satisfying $f_Q>0$.

The paper is organized as follows. 
In Sec.~\ref{sec:Symmetric_Teleparallel_theory} we briefly review the theoretical framework of STG and STEGR. 
Sec.~\ref{sec:Geodesic_and_its_deviation_equation} shows the geodesic equations and their deviation equations in Riemannian and symmetric teleparallel geometries. 
Sec.~\ref{sec:modified_Raychaudhuri_equation_in_f(Q)} concentrates on the definitions of the kinematic quantities in STG and presents the modified Raychaudhuri equation with matter contribution in $f(Q)$ gravity. In particular,
we impose a focusing condition in $f(Q)$ gravity to preserve the focusing theorem of the attractive behavior of gravity. 
Based on the result of the required focusing condition, 
we apply the FLRW background and deduce a focusing condition in Sec.~\ref{sec:Focusing_condition in flat FLRW}. 
This condition is then explicitly evaluated for the coincident gauge, 
which becomes $T_\text{eff}=T$.
Finally, conclusions are given in Sec.~\ref{sec:Conclusion}.

\section{\label{sec:Symmetric_Teleparallel_theory}Symmetric Teleparallel theory}
In teleparallel geometry (TG), the metric-affine connection $\Gamma^{\rho}{}_{\mu\nu}$ is postulated to be curvature-free, i.e., 
\begin{align}\label{eq:curvature-free}
    R_{\mu\nu\rho}{}^\sigma 
    = 2\partial_{[\mu}\Gamma^{\sigma}{}_{\nu]\rho} 
      + 2\Gamma^{\sigma}{}_{[\mu|\lambda|}\Gamma^{\lambda}{}_{\nu]\rho} = 0,
\end{align}
which is known as the teleparallel condition. Locally, a flat connection implies path-independent parallel transport. Along with the teleparallel condition $R_{\mu\nu\rho}{}^\sigma=0$, we also impose the torsion-free condition
\begin{align}\label{eq:torsion-free condition}
    T^\rho{}_{\mu\nu} = 2\Gamma^\rho{}_{[\mu\nu]} = 0.
\end{align}
The curvature-free and torsion-free conditions
reduce the geometry to the so-called symmetric teleparallel geometry (STG).  
The covariant derivative of a vector field $A^\rho$ is given by
\begin{equation}\label{eq:the covariant derivative of a vector field}    
{\nabla_{\!\mu}}{A}^\rho = \partial_{\mu}A^\rho+\Gamma^\rho{}_{\mu\nu}A^{\nu},
\end{equation}
where the STG connection can be decomposed into
\begin{align}\label{eq:connection_in_STG}
\Gamma^\rho{}_{\mu\nu}=\mathring{\Gamma}^\rho{}_{\mu\nu}+\mathrm{L}^\rho{}_{\mu\nu},
\end{align}
where $\mathring{\Gamma}^\rho{}_{\mu\nu}$ is the well-known Levi-Civita connection 
and $\mathrm{L}^\rho{}_{\mu\nu}$ is called the disformation tensor. 
In addition to the curvature and torsion tensors,
we define the nonmetricity tensor
\begin{align}\label{eq:nonmetricity tensor}
    Q_{\rho\mu\nu}
    :={\nabla_{\!\rho}}g_{\mu\nu}
    = \partial_{\rho}\,g_{\mu\nu}
      - \Gamma^{\sigma}{}_{\rho\mu}\,g_{\sigma\nu}
      - \Gamma^{\sigma}{}_{\rho\nu}\,g_{\mu\sigma},
\end{align}
which labels the failure of the connection to be metric-compatible. 
For a flat and torsion-free connection, one can always choose local coordinates \{$y^a$\} such that the STG connection $\Gamma^{a}{}_{bc}$ vanishes, 
i.e., we have $\Gamma^{a}{}_{bc}(y)=0$. This particular coordinate is referred to as the coincident gauge. Within this gauge, the fundamental variables of STG are the metric $g_{ab}$ and the coordinate functions $\{y^a\}$. 
However, in arbitrary coordinates \{$x^\mu$\}, the connection becomes 
\begin{align}\label{eq:metric-affine connection in arbitrary coordinates}
    \Gamma^{\rho}{}_{\mu\nu}(x)
    = \frac{\partial{x}^\rho}{\partial{y}^a}
      \frac{\partial }{\partial x^{\mu}}
      \frac{\partial y^a}{\partial x^{\nu}}.
\end{align}
Combining Eqs.~\eqref{eq:nonmetricity tensor} and \eqref{eq:connection_in_STG}, one can obtain the relation between the disformation tensor and the nonmetricity tensor.
\begin{align}\label{eq:L_from_Q}
    \mathrm{L}^\rho{}_{\mu\nu} 
    = \frac{1}{2}g^{\rho\alpha} 
      ( Q_{\alpha\mu\nu} - Q_{\mu\nu\alpha} - Q_{\nu\alpha\mu} ).
\end{align}

In STG, the curvature-free condition implies that the curvature vanishes, although the curvature associated with the Levi-Civita connection does not need to.
Hence, we obtain the relation 
\begin{align}\label{eq:STEGR relation}
\mathring{R}_{\mu\nu\rho}{}^{\sigma}
= - \mathring{\nabla}_{\!\mu}\,\mathrm{L}^{\sigma}{}_{\nu\rho}
  + \mathring{\nabla}_{\!\nu}\,\mathrm{L}^{\sigma}{}_{\mu\rho}
  - \mathrm{L}^{\sigma}{}_{\mu\lambda}\,\mathrm{L}^{\lambda}{}_{\nu\rho}
  + \mathrm{L}^{\sigma}{}_{\nu\lambda}\,\mathrm{L}^{\lambda}{}_{\mu\rho},
\end{align}
where $\mathring{\nabla}$ is the covariant derivative with respect to the Levi-Civita connection $\mathring{\Gamma}^\rho{}_{\mu\nu}$.
Consequently, the nonvanishing Ricci scalar $\mathring{R}$ defined by the Levi-Civita connection can be expressed by
\begin{align}\label{eq:STEGR relation 2}
\mathring{R} 
= Q - {\nabla_{\!\rho}}( Q^\rho 
  - \tilde{Q}^\rho ) - \frac{1}{2} Q _ \rho ( Q ^\rho 
  - \tilde{Q} ^ \rho )
\end{align}
where the scalar $Q$ 
is defined by 
\begin{align}\label{eq:Q_scalar}
    Q = Q_{\rho\mu\nu}P^{\rho\mu\nu}
\end{align}
with the so-called nonmetricity conjugate
\begin{align}\label{eq:nonmetricity_eff conjugate}
P^{\rho\mu\nu} 
:= - \frac{1}{4}Q^{\rho\mu\nu} 
   + \frac{1}{2}Q^{(\mu\nu)\rho} 
   + \frac{1}{4}(Q^\rho-\tilde{Q}^\rho)g^{\mu\nu} 
   - \frac{1}{4}g^{\rho(\mu}Q^{\nu)}.
\end{align}
Furthermore, two vectors have been defined through the contraction of the nonmetricity tensor:
$Q_\rho = Q_{\rho\mu\nu}g^{\mu\nu}$ 
and 
$\Tilde{Q}_\rho = Q_{\mu\nu\rho}g^{\mu\nu}$. 
Eq.~\eqref{eq:STEGR relation 2} shows that the difference between $Q$ and $\mathring{R}$ is a boundary term.
Therefore, 
we can use $Q$ as the Lagrangian density to construct an equivalent gravitational theory in STG, called STEGR.
Finally, 
the generalized gravitational action of $f(Q)$ gravity in STG can be written as
\begin{align}\label{eq:action_of_f(Q)}
    S = \frac{1}{2\kappa} \int \sqrt{-g}\, f(Q)\, \mathrm{d}^4x +S_m,
\end{align}
where $g := \det(g_{\mu\nu})$, $\kappa:=8 \pi G$ 
in geometric units of $G = c = 1$, and $S_m$ is the action of matter.
This is a generalization of the STEGR action. 
Varying the action \eqref{eq:action_of_f(Q)} with respect to the metric $g_{\mu\nu}$ should yield the field equation
\begin{align}\label{eq:EOM_in_STG}
   f_Q\mathring{G}^{\mu\nu} 
   + \frac{1}{2}( Qf_Q-f ) g^{\mu\nu} 
   + 2{\nabla_{\!\rho}}f_Q P^{\rho\mu\nu} 
   = \kappa T^{\mu\nu},
\end{align}
where $f_Q := \mathrm{d}f/\mathrm{d}Q$, $\mathring{G}^{\mu\nu}=\mathring{R}^{\mu\nu}-(1/2)\mathring{R}g^{\mu\nu}$ is the Einstein tensor associated with the Levi-Civita connection, 
and 
\begin{align}
T^{\mu\nu} := \frac{2}{\sqrt{-g}}\frac{\delta S _m}{\delta g _{\mu\nu}}
\end{align}
is the energy-momentum tensor of matter.
The corresponding trace equation of Eq.~\eqref{eq:EOM_in_STG} is
\begin{align}\label{eq:EOM_in_STG_trace}
\nabla_{\!\rho} \Big( f_Q (Q^\rho - \tilde{Q}^\rho ) \Big) \
+ f_Q \bigg( \frac{1}{2} Q_\rho ( Q^\rho - \tilde{Q} ^\rho ) - Q \bigg) + 2 ( Q f_Q - f ) = \kappa\, T,
\end{align}
where $T := T^{\mu\nu}{g}_{\mu\nu}$ is the trace of the energy-momentum tensor of the matter.
When $f(Q)=Q$, Eq.~\eqref{eq:EOM_in_STG} is reduced to the field equation of STEGR, which is dynamically equivalent to GR.

\section{\label{sec:Geodesic_and_its_deviation_equation}Geodesic and its deviation equation}
To study geodesics and their deviation, we first define the tangent vector field along the congruence
\begin{align}\label{eq:tangent vector field_0}
u^\rho := \frac{\mathrm{d}x^\rho}{\mathrm{d}\lambda},
\end{align}
where $\lambda$ is an affine parameter along each geodesic. 
Hereafter, we set $\lambda = \tau$ as the proper time along the timelike geodesic and simply use the normalization condition 
\begin{align}\label{eq:normalization condition of u}
g_{\mu\nu}u^{\mu}u^{\nu} = -1.
\end{align} 
The geodesic equation can always be obtained from the variation
of the proper time interval along a worldline between two events $A$ and $B$
\begin{align}\label{eq:proper time between two points}
    \delta \tau_{AB} 
    = \delta\int_{\tau_A}^{\tau_B}\sqrt{-g_{\mu\nu}u^{\mu}u^{\nu}}\,\mathrm{d}\tau,
\end{align}
which yields
\begin{align}\label{eq:geodesic}
\dot{u}^\rho + \mathring{\Gamma}^\rho{}_{\mu\nu}u^\mu{u}^\nu
=0.
\end{align}
The overdot symbol always denotes the derivative with respect to the proper time $\tau$ along the timelike geodesic. 
In the following, we compare the covariant form of the geodesics and their corresponding deviation equations in the Riemannian geometry and STG.

\subsection{\label{sec:Geodesic_and_its_deviation_equation in Riemann}Riemannian Geometry}
In Riemannian geometry, for any vector $V^{\rho}$, 
we can replace its ordinary derivative by the covariant derivative, i.e.,
\begin{subequations}\label{eq:procedure partial to nabla in Riemann}
\begin{align}\label{eq:procedure partial to nabla in Riemann 1}
\dot{V}^\rho 
= u^{\mu}\partial_{\mu} V^{\rho}
= u^{\mu}\mathring{\nabla}_{\!\mu} V^{\rho} 
  - \mathring{\Gamma}^{\rho}{}_{\mu\nu}u^{\mu}V^{\nu},
\end{align}
or
\begin{align}\label{eq:procedure partial to nabla in Riemann 2}
\dot{V}^\rho 
= \frac{\mathrm{d} V^{\rho}}{\mathrm{d}\tau}
= \frac{\mathring{\mathrm{D}} V^{\rho}}{\mathrm{d}\tau}
  - \mathring{\Gamma}^{\rho}{}_{\mu\nu}u^{\mu}V^{\nu},
\end{align}
\end{subequations}
where $\mathring{\mathrm{D}}/\mathrm{d}\tau$ is the absolute derivative associated with the Levi-Civita connection $\mathring{\Gamma}^\rho{}_{\mu\nu}$,
then the geodesic equation \eqref{eq:geodesic} has the covariant form of 
\begin{align}\label{eq:geodesic_in_Riem}
0 = \mathring{a}^{\rho} 
:= \frac{\mathring{\mathrm{D}} u^{\rho}}{\mathrm{d}\tau}
= u^\mu\mathring{\nabla}_\mu{u}^\rho,
\end{align}
where $\mathring{a}$ refers to the Riemannian acceleration.
This implies that the acceleration is defined through the covariant derivative and is equal to zero.
In addition, the autoparallel equation has the same form as 
Eq.~\eqref{eq:geodesic_in_Riem}, 
which implies that, in Riemannian geometry, the geodesic is both the shortest and the straightest curve in spacetime. 
The deviation vector between two adjacent geodesics is described by the geodesic deviation equation. 
Considering two events in two neighboring geodesics with the coordinates $x^\rho(\tau)$ and $x^\rho(\tau) + n^\rho(\tau)$, where $n^\rho$ is the deviation vector between two neighboring geodesics.
For the adjacent geodesic $x^\rho + n^\rho$, 
we should also have
\begin{align}\label{eq:geodesic_perturbation_equation}
(\dot{u}^\rho+\ddot{n}^\rho)+\mathring{\Gamma}^\rho{}_{\mu\nu}(x+n)(u^\mu+\dot{n}^\mu)(u^\nu+\dot{n}^\nu)=0.
\end{align}
Expanding the Levi-Civita connection to first order
\begin{align}\label{eq:expansion_of_connection}
    \mathring{\Gamma}^\rho{}_{\mu\nu}(x+n)\approx\mathring{\Gamma}^\rho{}_{\mu\nu}(x)+n^\alpha\partial_\alpha\mathring{\Gamma}^\rho{}_{\mu\nu},
\end{align}
and keeping terms linear in $n^{\rho}$ in Eq.~\eqref{eq:geodesic_perturbation_equation},
the difference between Eqs.~\eqref{eq:geodesic_perturbation_equation} and \eqref{eq:geodesic}
gives the relative acceleration between two geodesics,
which yields the linearized deviation equation
\begin{align}\label{eq:non_covariant_geodesic_deviation_equation_in_Riemann}
    \ddot{n}^\rho 
    + n^\alpha\,\partial_\alpha\mathring{\Gamma}^\rho{}_{\mu\nu}u^\mu u^\nu  
    + 2\mathring{\Gamma}^\rho{}_{\mu\nu}u^\mu\dot{n}^\nu=0.
\end{align}
Again, by using the language of the absolute derivative of Eq.~\eqref{eq:procedure partial to nabla in Riemann 2} with replacing
$u^{\rho}$ with $n^{\rho}$ or $\dot{n}^{\rho}$, 
one obtains the so-called equation of geodesic deviation in Riemannian geometry~\cite{levi1927ecart}
\begin{align}\label{eq:geodesic_deviation_equation_in_Riem}
\frac{\mathring{\mathrm{D}}^2 n^{\rho}}{\mathrm{d}\tau^2}
+ \mathring{R}_{\alpha\mu\nu}{}^\rho\, n^\alpha u^\mu u^\nu = 0.
\end{align}

\subsection{\label{sec:Geodesic_and_its_deviation_equation in STG}Symmetric Teleparallel Geometry}
The situation changes in STG
because the geodesic and autoparallel curves do not coincide with each other.
In the procedure of Eq.~\eqref{eq:procedure partial to nabla in Riemann},
we need to use the covariant derivative defined by the STG connection,
then Eq.~\eqref{eq:procedure partial to nabla in Riemann} should be modified as
\begin{subequations}\label{eq:procedure partial to nabla in STG}
\begin{align}\label{eq:procedure partial to nabla in STG 1}
\dot{V}^\rho 
= u^{\mu}\partial_{\mu} V^{\rho}
= u^{\mu}{\nabla_{\!\mu}} V^{\rho} 
  - \Gamma^{\rho}{}_{\mu\nu}u^{\mu}V^{\nu}
\end{align}
or
\begin{align}\label{eq:procedure partial to nabla in STG 2}
\dot{V}^\rho 
= \frac{\mathrm{d} V^{\rho}}{\mathrm{d}\tau}
= \frac{\mathrm{D} V^{\rho}}{\mathrm{d}\tau}
  - \Gamma^{\rho}{}_{\mu\nu}u^{\mu}V^{\nu},
\end{align}
\end{subequations}
where $\mathrm{D}/\mathrm{d}\tau=u^\alpha{\nabla_{\!\alpha}}$ is the absolute derivative along the geodesic.
Using Eqs.~\eqref{eq:connection_in_STG} and \eqref{eq:procedure partial to nabla in STG},
the geodesic equation \eqref{eq:geodesic} would become
\begin{align}\label{eq:geodesic_in_STG}
a^{\rho} := u^\mu{\nabla_{\!\mu}}{u}^\rho = \mathrm{L}^\rho{}_{\mu\nu}u^\mu{u}^\nu
\quad\text{or}\quad
a^{\rho} := \frac{\mathrm{D}{u}^{\rho}}{\mathrm{d}\tau} 
= \mathrm{L}^\rho{}_{\mu\nu}u^\mu{u}^\nu.
\end{align}
Since the geometry is not metric-compatible, lowering an index does not commute with covariant differentiation. 
Therefore, $u^\mu\nabla_{\!\mu} u_\rho$ is not equal to $a_\rho = a^\nu g_{\nu\rho}$ by the definition of the acceleration \eqref{eq:geodesic_in_STG}.
It should be
\begin{align}\label{eq:lowindex of acceleration}
    u^\mu \nabla_{\!\mu} u_\rho 
    = u^\mu \nabla_{\!\mu}(u^\nu g_{\nu\rho}) 
    = a_\rho + Q_{\mu\nu\rho} u^\mu u^\nu.
\end{align}
When the geodesic equation is rewritten using the STG connection, 
the non-vanishing acceleration is induced by the disformation tensor in Eq.~\eqref{eq:geodesic_in_STG} as a force-like term.
As a result, the geodesic equation does not coincide with the autoparallel equation
\begin{align}
u^\mu{\nabla_{\!\mu}}{u}^\rho=0
\quad\text{or}\quad
\frac{\mathrm{D}{u}^{\rho}}{\mathrm{d}\tau} = 0.
\end{align}

The linearized deviation equation in STG should still be the same as Eq.~\eqref{eq:non_covariant_geodesic_deviation_equation_in_Riemann}
because the metric geodesic equation remains determined by the Levi-Civita connection.
However, its physical interpretation should be formulated covariantly. 
The relative acceleration should be obtained through the acceleration defined by the covariant form of Eq.~\eqref{eq:geodesic_in_STG} in terms of the STG connection, i.e., 
the acceleration is defined by the STG connection rather than the Levi-Civita one. 
We must use the decomposition \eqref{eq:connection_in_STG} to implement the calculations.
Replacing 
$\partial_\alpha\mathring{\Gamma}^\rho{}_{\mu\nu}$
by
$\partial_\alpha\Gamma^\rho{}_{\mu\nu} - \partial_{\alpha}\mathrm{L}^{\rho}{}_{\mu\nu}$
in Eq.~\eqref{eq:non_covariant_geodesic_deviation_equation_in_Riemann}
and then using the curvature-free condition \eqref{eq:curvature-free} to rewrite 
$\partial_\alpha\Gamma^\rho{}_{\mu\nu}$, 
we obtain
\begin{align}\label{eq:non_covariant_geodesic_deviation_equation_in_STG}
\ddot{n}^\rho 
+ n^\alpha\big(
    \partial_{\mu}\Gamma^\rho{}_{\alpha\nu} 
    - \Gamma^{\rho}{}_{\alpha\lambda}\Gamma^{\lambda}{}_{\mu\nu} 
    + \Gamma^{\rho}{}_{\mu\lambda}\Gamma^{\lambda}{}_{\alpha\nu} 
    - \partial_{\alpha}\mathrm{L}^{\rho}{}_{\mu\nu}
  \big) u^\mu u^\nu  
+ 2\mathring{\Gamma}^\rho{}_{\mu\nu}u^\mu\dot{n}^\nu=0.
\end{align}
We also need to write 
$\partial_{\alpha}\mathrm{L}^{\rho}{}_{\mu\nu}$
in terms of 
${\nabla_{\!\alpha}}\mathrm{L}^{\rho}{}_{\mu\nu}$
and the STG connection.
Subsequently, using a similar procedure to write the partial derivative of 
$n^{\rho}$ or $\dot{n}^{\rho}$ 
in terms of its covariant/absolute derivative form, however, using Eq.~\eqref{eq:procedure partial to nabla in STG} rather than Eq.~\eqref{eq:procedure partial to nabla in Riemann},
and after some algebra, 
we eventually obtain the relative acceleration equation or
the covariant form of the geodesic deviation equation in STG 
\begin{align}\label{eq:covariant_geodesic_deviation_equation_in_STG}
\frac{\mathrm{D}^2 n^{\rho}}{\mathrm{d}\tau^2}
- 2\mathrm{L}^\rho{}_{\mu\alpha}u^\mu\,\frac{\mathrm{D}n^{\alpha}}{\mathrm{d}\tau}
- n^\alpha \big({\nabla_{\!\alpha}} \mathrm{L}^\rho{}_{\mu\nu}\big) u^\mu u^\nu = 0.
\end{align}
Eqs.~\eqref{eq:covariant_geodesic_deviation_equation_in_STG} and \eqref{eq:geodesic_deviation_equation_in_Riem} illustrate how geometry governs the evolution of geodesic deviations. 
In the Riemannian case, the relative acceleration is solely determined by the Riemann curvature tensor. 
In STG, however, the dynamics is different: 
both the disformation tensor $\mathrm{L}^\rho{}_{\mu\nu}$ and 
its covariant derivative ${\nabla_{\!\alpha}}\mathrm{L}^\rho{}_{\mu\nu}$ 
contribute to the evolution.
In particular, the former produces a damping-like term.
Therefore, the physical interpretation of the geodesic deviation equation differs from its Riemannian counterpart. 
It is inappropriate to define the relative acceleration by
Eq.~\eqref{eq:geodesic_deviation_equation_in_Riem} 
with a simple substitution of Eq.~\eqref{eq:STEGR relation}.
We note that the geodesic deviation equation \eqref{eq:covariant_geodesic_deviation_equation_in_STG}
 agrees with the result in Ref.~\cite{Capozziello:2024vix}.

\section{\label{sec:modified_Raychaudhuri_equation_in_f(Q)}Modified Raychaudhuri Equation in \texorpdfstring{$f(Q)$}{f(Q)} Gravity}
In this section, we first review the Raychaudhuri equation in GR and then derive the modified Raychaudhuri equation in $f(Q)$ gravity, 
which can be used to describe the evolution of a geodesic congruence. 
The coordinates of a spacetime event in a geodesic congruence are assumed to be parameterized by $x^\rho=x^\rho(\tau,s)$, 
where $\tau$ is the proper time along the timelike geodesic and $s$ labels different geodesics.
A velocity vector along the congruence and a deviation vector from one geodesic to a nearby one are
\begin{subequations}\label{eq:tangent vector fields}
\begin{align}\label{eq:velocity vector field}
u^\rho := \frac{\partial x^\rho}{\partial \tau}
\end{align}
and
\begin{align}\label{eq:deviation vector field}
\xi^\rho := \frac{\partial x^\rho}{\partial s},
\end{align}
\end{subequations}
respectively.
We first consider two simultaneous events located at 
$s$ and $s + \mathrm{d}s$ with fixed $\mathrm{d}s$.
Let $x^\rho(\tau,s)$ 
and 
$x^\rho(\tau,s + \mathrm{d}s)$ 
describe two neighboring geodesics passing through these two simultaneous events.
We note that, through the expansion of the parameter $s$, 
the coordinates of a neighboring geodesic separated by $\mathrm{d}s$ are given by
$x^\rho(\tau,s + \mathrm{d}s) = x^\rho(\tau,s) + \xi^\rho(\tau,s)\,\mathrm{d}s$, 
which induces the identification of the deviation vector with $\mathrm{d}s = 1$:
\begin{align}
n^{\rho}(\tau,s) 
\equiv x^\rho(\tau,s + 1) - x^\rho(\tau,s)
= \xi^\rho(\tau,s).
\end{align} 
By definitions of 
Eqs.~\eqref{eq:velocity vector field} and \eqref{eq:deviation vector field}, 
we obtain the equation of the Lie-dragging condition 
\begin{align}\label{eq:Lie-dragging condition}
\mathsterling_u\xi^\rho 
= [u,\xi]^\rho
= 0.
\end{align}
In addition, the covariant form of the Lie derivative can be written by
\begin{align}\label{eq:Lie_derivative} 
    \mathsterling_u\xi^\rho 
    = u^\alpha{\nabla_{\!\alpha}}\xi^\rho - \xi^\alpha{\nabla_{\!\alpha}}u^\rho,
\end{align}
through Eq.~\eqref{eq:Lie-dragging condition}, which yields
\begin{align}\label{eq:relation from Lie_derivative}
    \frac{\mathrm{D} \xi^\rho}{\mathrm{d}\tau} 
    = u^\alpha{\nabla_{\!\alpha}}\xi^\rho
    = \xi^\alpha{\nabla_{\!\alpha}}u^\rho,
\end{align}
which appears in the damping term of the geodesic deviation equation \eqref{eq:covariant_geodesic_deviation_equation_in_STG}.

\paragraph{Kinematic quantities}
Since the STG connection is not metric compatible, 
the covariant derivative of a covariant velocity
$u_{\mu} = g_{\mu\rho} u^{\rho}$ is not the same as the contravariant one
\begin{align}\label{eq:relation of convariant derivative of ud and uu}
{\nabla_{\!\nu}} u_{\mu}
= {\nabla_{\!\nu}} (g_{\mu\rho} u^{\rho}) 
= Q_{\nu\mu\rho} u^{\rho} + g_{\mu\rho} {\nabla_{\!\nu}} u^{\rho}.
\end{align}
We first introduce the so-called deformation tensor as a velocity gradient
\begin{align}\label{eq:bar-B tensor}
\bar{B}^{\mu}{}_{\nu} := {\nabla_{\!\nu}} u^\mu.
\end{align}
In addition, we have a similar tensor defined by 
$B_{\mu}{}_{\nu} := {\nabla_{\!\nu}} u_{\mu}$.
Consequently, the relation between these two rank-$2$ tensors of the $(2,0)$- and $(0,2)$-types should be
\begin{subequations}\label{eq:Bdd and Buu relation}
\begin{align}
B^{\mu\nu} 
&:= g^{\mu\alpha}g^{\nu\rho} B_{\alpha\rho}
= Q^{\nu\mu}{}_{\alpha}\,u^{\alpha}
  + \bar{B}^{\mu\nu}, \label{eq:Buu relation} \\
  B_{\mu}{}_{\nu} 
&= Q_{\nu\mu\alpha} u^{\alpha} + g_{\mu\alpha} \bar{B}^{\alpha}{}_{\nu}
:= Q_{\nu\mu\alpha} u^{\alpha} + \bar{B}_{\mu\nu}, \label{eq:Bdd relation}
\end{align} 
\end{subequations}
respectively.
In Ref.~\cite{Iosifidis:2018diy}, the authors use $B_{\mu}{}_{\nu}$ to define the kinematic quantities for calculations. 
Instead, we prefer to discuss $\bar{B}^{\mu}{}_{\nu}$ 
due to the nature of the subscript $\nu$ and the superscript $\mu$ of the covariant derivative ${\nabla_{\!\nu}}$ and the velocity vector $u^\mu$, respectively.
The tensor $\bar{B}^{\mu\nu} = g^{\nu\alpha}\bar{B}^{\mu}{}_{\alpha}$ is achieved by raising the index $\nu$ of the covariant derivative naturally without involving a nonmetricity tensor 
$Q_{\rho\mu\nu}$.
Correspondingly, the trace of $\bar{B}^{\mu\nu}$ defines a scalar
\begin{align}\label{eq:trace_part}
\bar{\theta} := \bar{B}^{\mu\nu}g_{\mu\nu}. 
\end{align}

As we have defined, $u^{\mu}$ is a tangent vector along a geodesic congruence,
a projection tensor onto the hypersurface orthogonal to $u^{\mu}$ can be defined by
\begin{align}\label{eq:projection}
h^{\mu\nu} := g^{\mu\nu}+u^\mu u^\nu,
\end{align}
which satisfies the condition 
$h^{\mu\nu}u_\mu=0$. 
One can define a purely spatial 
deformation tensor by projection through Eq.~\eqref{eq:projection}:
\begin{align}\label{eq:double_projected_tensor}
\mathcal{B}^{\mu\nu}
= h^\mu{}_{\alpha} h^{\nu\beta}\bar{B}^{\alpha}{}_{\beta}
\end{align}
with the orthogonality of 
$\mathcal{B}^{\mu\nu}u_\mu=0$ 
and
$\mathcal{B}_{\mu\nu}u^\nu=0$.
By expanding the right-hand side of Eq.~\eqref{eq:double_projected_tensor}, 
we can easily obtain the relations of the projected and unprojected deformation tensors of $(2,0)$- and $(0,2)$-types
\begin{subequations}\label{eq:expansion of projected deformation tensor}
\begin{align}
\mathcal{B}^{\mu \nu} 
&= \bar{B}^{\mu \nu} 
  + a ^{\mu} u^{\nu}
  + u^{\mu}g^{\nu\rho}(u_{\alpha}{\nabla_{\!\rho}}u^{\alpha})
  + u^{\mu}u^{\nu}u_{\alpha}a^{\alpha}, \label{eq:expansion of projected deformation tensor Buu}\\
\mathcal{B}_{\mu \nu} 
&= \bar{B}_{\mu \nu} 
  + a_{\mu} u_{\nu}
  + u_{\mu}(u_{\alpha}{\nabla_{\!\nu}}u^{\alpha})
  + u_{\mu}u_{\nu}u_{\alpha}a^{\alpha}. \label{eq:expansion of projected deformation tensor Bdd}
\end{align}
\end{subequations}
We recall the normalization condition \eqref{eq:normalization condition of u},
and its covariant derivative leads to an important result 
\begin{align}\label{eq:result of the normalization condition}
u_{\alpha}{\nabla_{\!\mu}}u^{\alpha} 
= -\frac{1}{2}Q_{\mu\alpha\beta}u^{\alpha}u^{\beta},
\end{align}
which is non-zero in STG.
It is obvious that the right-hand side is always zero in Riemannian geometry.
As a result, the relation of projected and unprojected deformation tensors is closely associated with Eq.~\eqref{eq:result of the normalization condition}.
Accordingly, the expansion scalar, shear tensor, and rotation tensor are defined by
\begin{subequations}\label{eq:expansion, rshear and rotation}
\begin{align}
    \theta
    &:= \mathcal{B}^{\mu\nu}h_{\mu\nu}
     = \bar{\theta} 
       + a^{\mu} u_{\mu}, \label{eq:expansion_scalar}\\[0.3em]
    \sigma^{\mu\nu}
    &:= \mathcal{B}^{(\mu\nu)} 
        - \frac{1}{3}\theta h^{\mu\nu}, \label{eq:shear_tensor}\\[0.3em]
    \omega^{\mu\nu}
    &:= \mathcal{B}^{[\mu\nu]}, \label{eq:rotation_tensor}
\end{align}
\end{subequations}
respectively,
which leads to the decomposition
\begin{align}
\mathcal{B}^{\mu \nu} 
= \sigma^{\mu\nu} + \omega^{\mu\nu} + \frac{1}{3}\theta h^{\mu\nu},
\end{align}
where 
$\mathcal{B}^{(\mu\nu)} = (1/2)(\mathcal{B}^{\mu\nu} + \mathcal{B}^{\nu\mu})$
and
$\mathcal{B}^{[\mu\nu]} = (1/2)(\mathcal{B}^{\mu\nu} - \mathcal{B}^{\nu\mu})$.
They should satisfy relations of
$\sigma^{\mu\nu}u_\nu=0$, 
$\omega^{\mu\nu}u_\nu=0$
and
$\sigma^\mu{}_{\mu}=0$.
Then, the corresponding algebraic decomposition of the deformation tensor $\bar{B}^{\mu \nu} $ becomes
\begin{align}\label{eq:decomposition_of_deformation_tensor}
\bar{B}^{\mu \nu} 
= \sigma^{\mu \nu} 
  + \omega^{\mu \nu} 
  - u^\nu a^\mu 
  - u_\alpha u^\mu \nabla^\nu u^\alpha 
  - u^\mu u^\nu u_\alpha a^\alpha 
  + \frac{1}{3} \theta h^{\mu \nu}.
\end{align}
We note that, by using Eq.~\eqref{eq:result of the normalization condition}, the second term on the right-hand side of Eq.~\eqref{eq:expansion_scalar} is non-vanishing unless the geometry is metric-compatible:
\begin{align}\label{eq:a dot u}
a^{\mu} u_{\mu} 
=  u^{\alpha}(u_{\mu}{\nabla_{\!\alpha}}u^{\mu}) 
= \bigg(-\frac{1}{2}u^{\alpha}Q_{\alpha\rho}{}^{\mu}u^{\rho}\bigg)u_{\mu} \neq 0,
\end{align}
which implies that the acceleration $a^{\mu}$ has a longitudinal component.
This explains why only $\theta$ is usually defined in the literature of metric theories without the term of $a^{\mu} u_{\mu}$ in Eq.~\eqref{eq:expansion_scalar}.

\paragraph{Hypersurface orthogonality}
By decomposition of the STG connection \eqref{eq:connection_in_STG},
the projected deformation tensor becomes
\begin{align}\label{eq:decomposition of Bud}
\mathcal{B}^{\mu}{}_{\nu} 
= \mathring{\mathcal{B}}^{\mu}{}_{\nu} 
  + h^{\mu}_{\alpha}\,h^{\beta}_{\nu}\,\mathrm{L}^{\alpha}{}_{\beta\gamma}u^{\gamma},
\end{align}
where $\mathring{\mathcal{B}}^{\mu}{}_{\nu} := \mathring{\nabla}_{\!\nu}u^{\mu}$.
The symmetric and antisymmetric parts in the first two indices of the disformation tensor \eqref{eq:L_from_Q} are read by
\begin{align}
\mathrm{L}_{(\mu\nu)\rho} &= -\frac{1}{2}Q_{\rho\mu\nu}, \\[0.3em]
\mathrm{L}_{[\mu\nu]\rho} &= Q_{[\mu\nu]\rho}.
\end{align}
Consequently, expansion, shear, and rotation have the forms 
\begin{subequations}\label{eq:decompositions of expansion, rshear and rotation in Riem}
\begin{align}
\theta
&= \mathring{\theta} - \frac{1}{2} h^{\alpha\beta}Q_{\gamma\alpha\beta}u^{\gamma}, \\[0.3em]
\sigma^{\mu\nu}
&= \mathring{\sigma}^{\mu\nu}
   - \frac{1}{2}h^{\mu\alpha}h^{\nu\beta}\,Q_{\gamma\alpha\beta}u^{\gamma}
   + \frac{1}{6}h^{\mu\nu}h^{\alpha\beta}Q_{\gamma\alpha\beta}u^{\gamma}, \label{eq:sigma under ho}\\[0.3em]
\omega^{\mu\nu} 
&= \mathring{\omega}^{\mu\nu} 
   + h^{\mu\alpha}h^{\nu\beta}\,Q_{[\alpha\beta]\gamma}u^{\gamma}, \label{eq:omega under ho}
\end{align}
\end{subequations}
where 
$\mathring{\theta} = \mathring{\mathcal{B}}^{\mu}{}_{\mu}$, 
$\mathring{\sigma}^{\mu\nu} 
:= \mathring{\mathcal{B}}^{(\mu\nu)} - (1/3)\mathring{\theta}h^{\mu\nu}$
and
$\mathring{\omega}^{\mu\nu} 
:= \mathring{\mathcal{B}}^{[\mu\nu]}$.
It is important to note that 
the hypersurface orthogonality does not lead to $\omega^{\mu\nu} = 0$.
By the Frobenius theorem in Riemannian geometry
\begin{align}
u_{[\mu}\mathring{\nabla}_{\!\nu}u_{\rho]} = 0, 
\end{align}
we have an identification of 
$u_{\alpha} = - N \partial_{\!\alpha}\phi$
for some scalar $\phi$ representing a hypersurface and lapse function $N$. 
Then it can be shown that the projected antisymmetric part of 
$\mathring{\mathcal{B}}_{\mu\nu}$ vanishes, i.e., $\mathring{\omega}_{\mu\nu} = 0$.
Therefore, the hypersurface orthogonality reduces the rotation tensor in STG to be
\begin{align}\label{eq:hypersurface orthogonality for omega}
\omega^{\mu\nu} = h^{\mu\alpha}h^{\nu\beta}Q_{[\alpha\beta]\gamma}u^{\gamma},
\end{align}
which does not vanish in general.
Consequently, the antisymmetric part of the nonmetricity $Q_{[\alpha\beta]\gamma}$ serves as the rotation effect of the geodesic congruence.

\subsection{General Relativity}\label{sec:GR}
In the Riemannian case, the curvature-free condition is relaxed, 
whereas the metricity condition is applied.
By setting the connection to be the Levi-Civita connection,
Eq.~\eqref{eq:relation from Lie_derivative} should be read as
\begin{align}\label{eq:relation from Lie_derivative in Riem}
\frac{\mathring{\mathrm{D}} \xi^\rho}{\mathrm{d}\tau} 
= \xi^\alpha \mathring{\nabla}_{\!\alpha}u^\rho,
\end{align}
and the deformation tensor is
\begin{align}
\mathring{B}_{\mu \nu} 
= \mathring{\nabla}_{\!\nu} u_{\mu}
= \mathring{\bar{B}}_{\mu \nu}.
\end{align}
The projected deformation tensor 
\eqref{eq:expansion of projected deformation tensor Buu} should be reduced to
$\mathring{\mathcal{B}}^{\mu \nu} = \mathring{B}^{\mu \nu}$ 
due to the reduction of 
$u_{\alpha}{\nabla_{\!\mu}}u^{\alpha} = 0$ in Eq.~\eqref{eq:result of the normalization condition}
and 
$a^{\mu} u_{\mu} = 0$ in Eq.~\eqref{eq:a dot u}.
Taking from Eqs.~\eqref{eq:decomposition of Bud}
and \eqref{eq:decompositions of expansion, rshear and rotation in Riem}, 
we obtain
$\theta = \mathring{\theta}$,
$\sigma^{\mu\nu} = \mathring{\sigma}^{\mu\nu}$
and 
$\omega^{\mu\nu} = \mathring{\omega}^{\mu\nu}$.

The second derivative of $\xi^\rho$ can be obtained from Eq.~\eqref{eq:relation from Lie_derivative in Riem}, it would yield the equation
\begin{align}
\frac{\mathring{\mathrm{D}}^2 \xi^\rho}{\mathrm{d}\tau^2}
= \xi^\alpha\Bigg( \mathring{B}^{\rho}{}_{\beta}\mathring{B}^{\beta}{}_{\alpha}
  + \frac{\mathring{\mathrm{D}} }{\mathrm{d}\tau} \mathring{B}^{\rho}{}_{\alpha}\Bigg)
\end{align}
Replacing the left-hand side of the previous equation by Eq.~\eqref{eq:geodesic_deviation_equation_in_Riem} with $n^{\alpha} = \xi^{\alpha}$ leads to
\begin{align}
\frac{\mathring{\mathrm{D}} }{\mathrm{d}\tau} \mathring{B}^{\rho}{}_{\alpha}
+ \mathring{B}^{\rho}{}_{\beta}\mathring{B}^{\beta}{}_{\alpha}
= - \mathring{R}_{\alpha\mu\nu}{}^{\rho}\, u^\mu u^\nu.
\end{align}
The trace of the resultant equation should be
\begin{align}
\frac{\mathring{\mathrm{D}} \mathring{\theta}}{\mathrm{d}\tau} 
+ \mathring{B}^{\alpha}{}_{\beta}\mathring{B}^{\beta}{}_{\alpha}
= - \mathring{R}_{\mu\nu}\, u^\mu u^\nu.
\end{align}
It is easy to obtain the second term by direct calculations:
\begin{align}
\mathring{B}^{\alpha}{}_{\beta}\mathring{B}^{\beta}{}_{\alpha}
= -2\big( \mathring{\omega}^2 - \mathring{\sigma}^2
\big) 
  + \frac{1}{3}\mathring{\theta}^2,
\end{align}
where 
$2\,\mathring{\omega}^2 
:= \mathring{\omega}^{\alpha\beta}\mathring{\omega}^{\mu\nu}g_{\alpha\mu}g_{\beta\nu}$ 
and 
$2\,\mathring{\sigma}^2 
:= \mathring{\sigma}^{\alpha\beta}\mathring{\sigma}^{\mu\nu}g_{\alpha\mu}g_{\beta\nu}$.
According to the Einstein field equations
\begin{align}\label{eq:Eom_of_GR}
\mathring{R}_{\mu\nu} = \kappa \bigg(T_{\mu\nu} -\frac{1}{2}g_{\mu\nu}T\bigg),
\end{align}
the standard Raychaudhuri equation in GR is 
\begin{align}\label{eq:Raychaudhuri_equation_in_GR_1}
\frac{\mathrm{D}\mathring{\theta}}{\mathrm{d}\tau}
= 2\big(\mathring{\omega}^2-\mathring{\sigma}^2\big) 
  - \frac{1}{3}\mathring{\theta}^2 
  - \kappa\bigg(T_{\mu\nu} - \frac{1}{2}g_{\mu\nu}T\bigg) u^\mu{u}^\nu.
\end{align}

The focusing condition and its formulation in GR can be briefly recalled below.
In the context of singularity theorems, we consider irrotational congruences due to hypersurface orthogonality, i.e., $\mathring{\omega}_{\mu\nu} = 0$ for metric theory, which simplifies the Raychaudhuri equation to
\begin{align}\label{eq:singularity_theorems_of_GR_1}
\frac{\mathrm{D}\mathring{\theta}}{\mathrm{d}\tau} 
= -2\mathring{\sigma}^2 - \frac{1}{3}\mathring{\theta}^2 
  - \kappa\bigg(T_{\mu\nu} - \frac{1}{2}g_{\mu\nu}T\bigg) u^{\mu}u^{\nu}.
\end{align}
Imposing the strong energy condition~\cite{Hawking:1966vg} 
\begin{align}\label{eq:strong_energy_eff condition_in_GR}
\bigg(T_{\mu\nu} - \frac{1}{2}g_{\mu\nu}T\bigg)u^\mu{u}^\nu\geq0,
\end{align}
This condition ensures attractive gravity for timelike geodesics and is satisfied by many ordinary matter models, which leads to
\begin{align}\label{eq:singularity_theorems_of_GR_2}
\frac{\mathrm{D}\mathring{\theta}}{\mathrm{d}\tau}
\leq 
- 2\mathring{\sigma}^2-\frac{1}{3}\mathring{\theta}^2.
\end{align}
Since the shear term $-2\mathring{\sigma}^2$ is always non-positive and promotes focusing, we drop it to obtain a weaker but still sufficient bound:
\begin{align}\label{eq:singularity_theorems_of_GR_3}
\frac{\mathrm{D}\mathring{\theta}}{\mathrm{d}\tau}\leq-\frac{1}{3}\mathring{\theta}^2.
\end{align}
This inequality implies that geodesic focusing is inevitable along the evolution of the geodesic congruence. In particular, if the geodesic congruence is initially converging, i.e., $\mathring{\theta}_0 < 0$, then the expansion diverges to $-\infty$ in finite proper time.

\subsection{\label{sec:f(Q) gravity}\texorpdfstring{$f(Q)$}{f(Q)} Gravity}
Taking the absolute derivative on Eq.~\eqref{eq:relation from Lie_derivative} with respect to $\tau$ again, 
we obtain
\begin{align}\label{eq:D^2 xi}
\frac{\mathrm{D^2} \xi^\rho}{\mathrm{d}\tau^2} 
= \frac{\mathrm{D}\xi^\alpha}{\mathrm{d}\tau}\bar{B}^\rho{}_{\alpha}
  + \xi^\alpha\frac{\mathrm{D}\bar{B}^\rho{}_{\alpha}}{\mathrm{d}\tau} 
= \xi^\alpha \Bigg(\bar{B}^\rho{}_{\mu}\bar{B}^\mu{}_{\alpha}
  + \frac{\mathrm{D}\bar{B}^\rho{}_{\alpha}}{\mathrm{d}\tau}\Bigg).
\end{align}
The left-hand side of Eq.~\eqref{eq:D^2 xi} can be substituted by the geodesic deviation equation \eqref{eq:covariant_geodesic_deviation_equation_in_STG},
and from Eq.~\eqref{eq:relation from Lie_derivative},
we obtain
\begin{align}\label{eq:6}
\xi^\alpha\Bigg(
  \frac{\mathrm{D}\bar{B}^\rho{}_{\alpha}}{\mathrm{d}\tau} 
  + \bar{B}^\rho{}_{\mu}\bar{B}^\mu{}_{\alpha}
  - {\nabla_{\!\alpha}}\mathrm{L}^\rho{}_{\mu\nu}u^\mu u^\nu
  - 2\mathrm{L}^\rho{}_{\mu\nu}u^\mu \bar{B}^\nu{}_{\alpha} \Bigg)
= 0.
\end{align}
For arbitrary $\xi^\alpha \neq 0$,
we should obtain the evolution equation 
\begin{align}\label{eq:7}
    \frac{\mathrm{D}\bar{B}^\rho{}_{\alpha}}{\mathrm{d}\tau}+\bar{B}^\rho{}_{\mu}\bar{B}^\mu{}_{\alpha}-{\nabla_{\!\alpha}}\mathrm{L}^\rho{}_{\mu\nu}u^\mu u^\nu-2\mathrm{L}^\rho{}_{\mu\nu}u^\mu \bar{B}^\nu{}_{\alpha}=0.
\end{align}
The last two terms on the left-hand side of Eq.~\eqref{eq:7} can be combined into one term
\begin{align}\label{eq:8}
    {\nabla_{\!\alpha}}\mathrm{L}^\rho{}_{\mu\nu}u^\mu u^\nu+2\mathrm{L}^\rho{}_{\mu\nu}u^\mu {\nabla_{\!\alpha}} u^\nu={\nabla_{\!\alpha}}(\mathrm{L}^\rho{}_{\mu\nu}u^\mu u^\nu).
\end{align}
With the definition of acceleration \eqref{eq:geodesic_in_STG},
Eq.~\eqref{eq:7} should be
\begin{align}\label{eq:The_intermediate_steps_of_Raychaudhuri_equation}
    \frac{\mathrm{D}\bar{B}^\rho{}_{\alpha}}{\mathrm{d}\tau}+\bar{B}^\rho{}_{\mu}\bar{B}^\mu{}_{\alpha}-{\nabla_{\!\alpha}} a^\rho=0.
\end{align}
We also provide the derivation of Eq.~\eqref{eq:The_intermediate_steps_of_Raychaudhuri_equation}
by passing the geodesic deviation equation \eqref{eq:covariant_geodesic_deviation_equation_in_STG} as a cross check in Appendix~\ref{app:Alternative derivation}.
Substituting 
Eq.~\eqref{eq:decomposition_of_deformation_tensor} into 
Eq.~\eqref{eq:The_intermediate_steps_of_Raychaudhuri_equation} 
and then contracting with $\delta_\rho^\alpha$ yields the equation 
\begin{align}\label{eq:modified_Raychaudhuri_equation_in_STG_0}
    \frac{\mathrm{D}\bar{\theta}}{\mathrm{d}\tau}
    = 2(\omega^2 - \sigma^2) 
      - \frac{1}{3}\theta^2
      +(a^\alpha u _ \alpha)^2
      +2u_\alpha a^\mu\nabla_\mu u^\alpha
      + {\nabla_{\!\rho}} a^\rho.
\end{align} 
This is the evolution equation of the unprojected scalar $\bar{\theta}$. 

The more fundamental consequence of the geodesic equation can be found 
if we take the covariant derivative of 
the normalization condition \eqref{eq:normalization condition of u}
along the congruence
which yields
 \begin{align}\label{eq:Simplification_condition_0}
     u_\rho A^\rho=0.
 \end{align}
 We have defined
 \begin{align}\label{eq:Definition_of_A}
 A^\rho
 := g^{\rho\mu}u^\nu u^\sigma Q_{\sigma\mu\nu} 
   + 2\mathrm{L}^\rho{}_{\mu\nu}u^\mu u^\nu,
 \end{align}
 where $A^\rho$ measures the mismatch between the metric variation along the congruence and the disformation-induced acceleration, so that Eq.~\eqref{eq:Simplification_condition_0} can be written as $u_\rho A^\rho=0$, which implies that $A^\rho$ is orthogonal to $u^\rho$. In what follows, we further restrict our analysis to the special case of
\begin{align}\label{eq:A = 0}
A^\rho=0,
\end{align} 
which implies that the metric variation along the congruence and the disformation-induced acceleration cancel each other out completely. 
This condition selects a restricted class of congruences and nonmetricity configurations in which the metric variation along the congruence is exactly converted into a disformation-induced acceleration term,
which we can call the nonmetricity-acceleration conversion equation.
Therefore, the subsequent results in this subsection are derived under this assumption.
Under this assumption, we obtain 
\begin{align}\label{eq:Simplification_condition_1}
    \mathrm{L}^\rho{}_{\mu\nu}u^\mu u^\nu
    = -\frac{1}{2}g^{\rho\mu}u^\nu u^\sigma Q_{\sigma\mu\nu}.
\end{align}
As a result, the geodesic equation \eqref{eq:geodesic_in_STG} should be rewritten as
\begin{align}\label{eq:geodesic_imposing_an_additional_condition}
    a^{\rho}
    = u^\mu{\nabla_{\!\mu}} u^\rho 
    = -\frac{1}{2}g^{\rho\mu}u^\nu u^\sigma Q_{\sigma\mu\nu}.
\end{align}
It is explicit that in STG, 
the departure of a geodesic from an autoparallel of the STG connection is governed by the nonmetricity tensor $Q_{\sigma\mu\nu}$.
In the metric-compatible limit of 
$Q_{\sigma\mu\nu} = 0$, 
the right-hand side of Eq.~\eqref{eq:geodesic_imposing_an_additional_condition} vanishes, 
namely the autoparallel equation $u^\mu{\nabla_{\!\mu}} u^\rho=0$.
In this limit, the geodesic coincides with the autoparallel curve with a flat connection in Minkowski space.

Furthermore, we impose the Weyl-type nonmetricity for simplicity, i.e.,
\begin{subequations}\label{eq:Weyl-type ansatz}
\begin{align}\label{eq:Weyl-type_nonmetricity_tensor}
    {\nabla_{\!\rho}}g_{\mu\nu}
    = Q_{\rho\mu\nu}
    = \varphi_{\rho}\,g_{\mu\nu},
\end{align}
which describes a conformal rescaling of the metric 
under the parallel transport.
The vector 
\begin{align}
\varphi_{\rho} = \frac{1}{4}\, Q_{\rho}.
\end{align}
\end{subequations}
can be easily obtained by taking a contraction with $g^{\mu\nu}$.
Within the Weyl-type ansatz, we have the following equations
\begin{subequations}\label{eq:absolute_derivative_of_metric_in_weyl_form}
\begin{align}
u^\alpha{\nabla_{\!\alpha}} g_{\mu\nu}
&= \frac{1}{4}u^\alpha{Q}_\alpha{g}_{\mu\nu}, \label{eq:absolute_derivative_of_metric_in_weyl_form_1} \\[0.3em]
u^\alpha{\nabla_{\!\alpha}} g^{\mu\nu}
&= -\frac{1}{4}u^\alpha{Q}_\alpha{g}^{\mu\nu}, \label{eq:absolute_derivative_of_metric_in_weyl_form_2} \\[0.3em]
\tilde{Q}_{\rho} 
&= \frac{1}{4}Q_{\rho}, \label{eq:absolute_derivative_of_metric_in_weyl_form_3} \\[0.3em]
P^{\rho\mu\nu}
&= \frac{1}{8}Q^{\rho}g^{\mu\nu}
   - \frac{1}{16}Q^{\mu}g^{\nu\rho}
   - \frac{1}{16}Q^{\nu}g^{\mu\rho}. \label{eq:absolute_derivative_of_metric_in_weyl_form_4} \\[0.3em]
Q
&= \frac{3}{32}Q_{\rho}Q^{\rho}, \label{eq:absolute_derivative_of_metric_in_weyl_form_5} 
\end{align}
\end{subequations}
and the acceleration \eqref{eq:geodesic_imposing_an_additional_condition} takes the form of
\begin{align}\label{eq:geodesic_in_STG_under_ansatz}
    a^{\rho} = -\frac{1}{8} u^\rho u^\alpha Q_\alpha,
\end{align}
which implies that the acceleration should be purely longitudinal.
Multiplying $u_{\rho}$ on both sides of the 
acceleration equation \eqref{eq:geodesic_in_STG_under_ansatz}, 
we obtain
\begin{align}\label{eq:inner product of u and a}
u_\rho a^\rho = \frac{1}{8} u^\alpha Q_\alpha.
\end{align}
On the other hand,
Eq.~\eqref{eq:L_from_Q} is read as  
\begin{align}\label{eq:Luu_2}
a^{\rho} 
= -\frac{1}{8}\big(Q^\rho+2u^\rho u^\alpha Q_\alpha\big),
\end{align}
which should be equal to Eq.~\eqref{eq:geodesic_in_STG_under_ansatz},
and then we obtain an identity
\begin{align}\label{eq:stronger_constraint_between_Q_and_u}
    Q^\rho = -u^\rho u^\alpha Q_\alpha
\end{align}
under the Weyl-type ansatz with the nonmetricity-acceleration conversion equation $A^\rho = 0$.
Therefore, 
the nonmetricity vector $Q^\rho$ is either zero or timelike and parallel/antiparallel to the congruence tangent.
By contracting Eq.~\eqref{eq:stronger_constraint_between_Q_and_u} with $Q_\rho$, the new identity is 
\begin{align}\label{eq:scalar_constraint_between_Q_and_u}
\big(u^\alpha Q_\alpha\big)^2 
= - Q^\alpha Q_\alpha
= - \frac{32}{3}\,Q.
\end{align}
It is manifest that Eq.~\eqref{eq:stronger_constraint_between_Q_and_u}
reduces the acceleration 
\eqref{eq:geodesic_in_STG_under_ansatz} to be
\begin{align}\label{eq:geodesic_in_STG_under_ansatz_2}
    a^{\rho} = \frac{1}{8}\,Q^{\rho},
\end{align}
and Eq.~\eqref{eq:modified_Raychaudhuri_equation_in_STG_0} should become
\begin{align}\label{eq:modified_Raychaudhuri_equation_in_STG_1}
    \frac{\mathrm{D}\bar{\theta}}{\mathrm{d}\tau}
    = 2(\omega^2 - \sigma^2) 
      - \frac{1}{3}\theta^2
      +\frac{1}{6}Q
      + \frac{1}{8} {\nabla_{\!\rho}} Q^\rho.
\end{align}
On account of Eq.~\eqref{eq:absolute_derivative_of_metric_in_weyl_form},
the trace of the field equation \eqref{eq:EOM_in_STG_trace} 
can be written as 
\begin{align}\label{eq:expression_of_substitution}
{\nabla_{\!\rho}} Q^\rho 
= \frac{4\kappa T}{3f_Q} 
  - Q^\rho{\nabla_{\!\rho}}\ln|f_Q| 
  - 4\,Q
  - \frac{8}{3}\bigg(Q - \frac{f}{f_Q}\bigg),
\end{align}
which appears in the last term on the right-hand side of Eq.~\eqref{eq:modified_Raychaudhuri_equation_in_STG_1}.
This implies that the modified Raychaudhuri equation is associated with matter sources and is not only a geometric relation any more.
In addition, according to the discussion in Ref.~\cite{BeltranJimenez:2019tme}, 
the effective Newton's constant in $f(Q)$ gravity can be defined by $G/f_{Q}$, 
therefore, $f_Q>0$ is required.
Finally, the modified Raychaudhuri equation should be the evolution equation of $\theta$ by using the relation of Eq.~\eqref{eq:expansion_scalar}: 
\begin{align}\label{eq:modified_Raychaudhuri_equation_in_STG_f(Q)}
\frac{\mathrm{D}\theta}{\mathrm{d}\tau}
&= \frac{\mathrm{D}\bar{\theta}}{\mathrm{d}\tau}
   + \frac{1}{8} \frac{\mathrm{D} }{\mathrm{d}\tau}\big(u^\alpha Q_\alpha\big).
\end{align}
The derivative in last term can be written as
\begin{align}
\frac{\mathrm{D}}{\mathrm{d}\tau}\big(u^{\alpha} Q_{\alpha}\big)
 = u^{\nu} \big({\nabla_{\!\nu}}u_{\alpha}\big) Q^{\alpha}
   + u^{\nu} u_{\alpha} 
     \big({\nabla_{\!\nu}} Q^{\alpha}\big).
\end{align}
Using Eq.~\eqref{eq:lowindex of acceleration}, the first term on the right-hand side can be expressed as
\begin{align}
u^{\nu} \big({\nabla_{\!\nu}}u_{\alpha}\big) Q^{\alpha} = -\frac{4}{3}\,Q.
\end{align}
We finally obtain the modified Raychaudhuri equation in $f(Q)$ gravity:
\begin{align}\label{eq:modified_Raychaudhuri_equation_in_STG_f(Q)_2}
\frac{\mathrm{D}\theta}{\mathrm{d}\tau}
&= 2(\omega^2-\sigma^2)
   - \frac{1}{3}\theta^2
   + \frac{\kappa T}{6f_Q} 
   - \frac{1}{8}Q^\rho{\nabla_{\!\rho}}\ln|f_Q|  \nonumber \\[0.3em]
&\qquad  - \frac{1}{2}\,Q
   - \frac{1}{3}\bigg(Q - \frac{f}{f_Q}\bigg)
   + \frac{1}{8}u^{\nu} u_{\alpha} \big({\nabla_{\!\nu}} Q^{\alpha}\big).
\end{align}
However, the focusing theorem is more subtle in $f(Q)$ gravity, 
this will be discussed in the next subsection.

\subsection{\label{subsec:Focusing_condition_of f(Q) gravity}Focusing Condition of \texorpdfstring{$f(Q)$}{f(Q)} Gravity}
The focusing theorem in GR has been recalled in Sec.~\ref{sec:GR}.
In this subsection,
we would like to concentrate on 
the non-trivial behavior of 
$\mathrm{D}\theta/\mathrm{d}\tau$ 
in $f(Q)$ gravity.
Under the Weyl-type ansatz,
we have 
$\sigma^{\mu\nu} = \mathring{\sigma}^{\mu\nu}$ and $\omega^{\mu\nu} = 0$ as in the Riemannian case for hypersurface orthogonality,
which can be found in Appendix~\ref{app:Hypersurface Orthogonality with Weyl-type Ansatz}.
Alternatively, one may restrict attention to an irrotational geodesic congruence for which $\omega^{\mu\nu} = 0$, 
accordingly, the rotational contribution vanishes.
We define an auxiliary function by
\begin{align}\label{eq:new_auxiliary_quantity}
F(Q,f_Q,T,\dots)
&= - 2\sigma^2 
   + \frac{\kappa T}{6f_Q} 
   - \frac{1}{8}Q^\rho{\nabla_{\!\rho}}\ln|f_Q| \nonumber \\[0.3em]
&\quad 
   - \frac{1}{2}Q
   - \frac{1}{3}\bigg( Q - \frac{f}{f_Q} \bigg)
   + \frac{1}{8}u^{\nu} u_{\alpha} \big({\nabla_{\!\nu}} Q^{\alpha}\big).
\end{align}
such that the modified Raychaudhuri equation takes the form
\begin{align}\label{eq:restricted_Raychaudhuri_0}
\frac{\mathrm{D}\theta}{\mathrm{d}\tau}
= F(Q,f_Q,T,\dots) -\frac{1}{3}\theta^2.
\end{align}
Except for the term $-(1/3)\theta^2$, the sign of the function $F(Q,f_Q,T,\dots)$ should be non-positive to satisfy the focusing theorem. 
Clearly, we have to implement a focusing condition:
\begin{align}\label{eq:FC_0}
    F(Q,f_Q,T,\dots) \leq 0.
\end{align}
Isolating the trace of the energy-momentum tensor, this condition can be rewritten as
\begin{align}\label{eq:FC_1}
    T \leq {T}_\text{eff}
\end{align}
due to $f_Q>0$,
where we have defined the effective energy-momentum trace
\begin{align}\label{eq:effective_trace}
{T}_\text{eff}
= \frac{6 f_Q}{\kappa}\Bigg[
    2\sigma^2 
    + \frac{1}{8}Q^\rho{\nabla_{\!\rho}}\ln|f_Q| 
    + \frac{1}{2}Q
    + \frac{1}{3}\bigg(Q-\frac{f}{f_Q}\bigg) 
    - \frac{1}{8}u^{\nu} u_{\alpha} \big({\nabla_{\!\nu}} Q^{\alpha}\big)
    \Bigg].
\end{align}
If $T\leq T_\text{eff}$, for an initially converging timelike congruence, $\theta$ diverges to $-\infty$ in finite proper time.
Otherwise, the nonmetricity effects may hinder the geodesic convergence, providing an additional mechanism to avoid singularities in certain $f(Q)$ models.

\section{\label{sec:Focusing_condition in flat FLRW}Focusing Condition in a Flat FLRW Background}
We consider the flat FLRW background as an example to examine the focusing condition in $f(Q)$ gravity. 
The metric in conformal time $\eta$ is given by
\begin{align}\label{eq:FLRW_metric}
\mathrm{d}s^2 = 
a^2(\eta) (-\mathrm{d}\eta^2+\delta_{ij}\mathrm{d}x^i\mathrm{d}x^j),
\end{align}
where $a(\eta)$ is the scale factor 
and $\delta_{ij}$ is the three-dimensional Euclidean metric tensor.
Under these coordinates, the four-velocity defined by the conformal time should become 
$u^{\rho} = (1/a,0,0,0)$.
Due to the cosmological principle of homogeneity and isotropy, 
the comoving timelike congruence in the flat FLRW metric should be shear-free and rotation-free:
\begin{align}
    \sigma^{\mu\nu}=0
    \quad\text{and}\quad
    \omega^{\mu\nu}=0,
\end{align}
respectively.

The flat FLRW metric is homogeneous and isotropic. 
Mathematically, 
it can be realized by
$\mathsterling_\chi{g}_{\mu\nu}=0$
in Riemannian geometry, 
in which the homogeneity and isotropy of the flat three-space is generated by
the six Killing vector fields $\chi^\rho$.
In STG we further require the connection $\Gamma^\rho{}_{\mu\nu}$ to satisfy the homogeneity and isotropy of cosmological symmetry~\cite{Hohmann:2020zre, Hohmann:2021ast, Heisenberg:2022mbo}
\begin{align}\label{eq:homogeneous_and_isotropic}
 \mathsterling_\chi\Gamma^\rho{}_{\mu\nu}
 = {\nabla_{\!\mu}}{\nabla_{\!\nu}}\chi^\rho = 0.
\end{align}
The non-zero components of the STG connection are consistent with the homogeneity and 
isotropy conditions \eqref{eq:homogeneous_and_isotropic} given by
\begin{align}\label{eq:non-zero_components}
\Gamma^0{}_{00} = K_1,
\quad
\Gamma^0{}_{ij} = K_2\, \delta_{ij},
\quad
\Gamma^i{}_{0j} = \Gamma^i{}_{j0} = K_3\, \delta_{ij},
\end{align}
where $\{K_1, K_2, K_3 \}$ 
are functions of conformal time $\eta$ and $i,j = 1,2,3$ denote spatial indices. 
The solutions for these functions can be divided into three branches 
\begin{subequations}
\begin{align}
&\text{Branch 1}:
\quad
K_1 = \gamma_1, 
\quad
K_2 = 0, 
\quad
K_3 = 0, \label{eq:branch_1} \\[0.3em]
&\text{Branch 2}:
\quad
K_1 = \frac{\gamma^{\,\prime}_2}{\gamma_2} + \gamma_2, 
\quad 
K_2 = 0, 
\quad 
K_3 = \gamma_2, \label{eq:branch_2} \\[0.3em]
&\text{Branch 3}:
\quad
K_1 = -\frac{\gamma^{\,\prime}_3}{\gamma_3}, 
\quad 
K_2 = \gamma_3, 
\quad 
K_3 = 0,\label{eq:branch_3}
\end{align}
\end{subequations}
where $\gamma_1, \gamma_2$, and $\gamma_3$ are arbitrary functions of conformal time and 
the prime denotes the derivative with respect to conformal time $\eta$. 
Only Branch 1 with $\gamma_1 = 0$ corresponds to the coincident gauge in the FLRW background.

Within the background \eqref{eq:FLRW_metric} and the 
connection given in Eq.~\eqref{eq:non-zero_components}, 
we have
\begin{align}
    Q = -\frac{3\big[2\mathcal{H}^2+K_1(K_2-K_3)+K_3^2-2\mathcal{H}(K_2+K_3)\big]}{a^2},
\end{align}
and the corresponding modified Friedmann equations of $f(Q)$ gravity with a single-component perfect fluid are
\begin{subequations}\label{eq:modified Friedmann equations}
\begin{align}
    3f_Q\mathcal{H}^2+\frac{1}{2}a^2(f-Qf_Q)+\frac{3}{2}(K_3-K_2)f'_Q &= \kappa a^2\rho,\\[0.3em]
    f_Q(2\mathcal{H}'+\mathcal{H}^2)+\frac{1}{2}a^2(f-Qf_Q)+\frac{1}{2}(4\mathcal{H}-K_2-3K_3)f'_Q &= -\kappa a^2p,
\end{align}
\end{subequations}
where $\rho$, $p$ and $\mathcal{H}=a'/a$ are the energy density, pressure, and Hubble parameter, respectively. 
The trace of energy-momentum tensor for a single-component fluid is $T = -\rho + 3 p$.

For the homogeneous and isotropic STG connection in the flat FLRW universe, the purely spatial components of the
nonmetricity tensor $Q _ {ijk}$ vanish, i.e.,
\begin{align}
    Q _ {ijk} = 0.
\end{align}
On the other hand, the Weyl-type ansatz requires that
\begin{align}
    Q _ {ijk} = a ^ 2 \varphi _ i \delta _ {jk}. 
\end{align}
Therefore, the compatibility between the Weyl-type ansatz and the homogeneous and isotropic STG connection requires
\begin{align}
    \varphi _ i = 0. 
\end{align}
All the non-zero components of the nonmetricity tensor should be
\begin{subequations}\label{eq:mixed temporal-spatial components 1}
\begin{align}
Q _ {000} &= 2 a ^ 2 (K_1 - \mathcal{H}),\\[0.3em]
Q _ {0ij} &= 2 a ^ 2 (\mathcal{H} - K_3) \delta _ {ij},\\[0.3em]
Q _ {i0j} &= Q _ {ij0} = a ^ 2 (K_2 - K_3) \delta _ {ij}.
\end{align}
\end{subequations}
On the other hand, the corresponding components of the nonmetricity tensor with the Weyl-type ansatz are
\begin{subequations}\label{eq:mixed temporal-spatial components 2}
\begin{align}
Q _ {000} &= - a ^ 2 \varphi_{0},\\[0.3em]
Q _ {0ij} &= a ^ 2 \varphi_{0} \delta _ {ij},\\[0.3em]
Q _ {i0j} &= Q _ {ij0} = 0.
\end{align}
\end{subequations}
Comparing Eqs.~\eqref{eq:mixed temporal-spatial components 1} 
and \eqref{eq:mixed temporal-spatial components 2} yields 
the condition
$K _ 1 = K _ 2 = K _ 3$. 
Therefore, among the three branches, Branches 2 and 3 admit no regular configurations compatible with the Weyl-type ansatz, whereas Branch 1 is compatible only when $K_1 = \gamma_1 = 0$, corresponding to the coincident gauge.
The effective trace $T_\text{eff}$ can be expressed as
\begin{align}\label{eq:effective trace in FLRW universe}
  {T}_\text{eff} = - \frac{2}{\kappa a ^ 2} \big( 3 f_Q ( \mathcal{H} ^ 2 + \mathcal{H}' ) + 3 \mathcal{H} f _ Q' +  a ^ 2 (f-Qf_Q) \big).
\end{align}
Using modified Friedmann equations \eqref{eq:modified Friedmann equations},
the effective trace should be
\begin{align}
    {T}_\text{eff}= -\rho + 3p,
\end{align}
which is equal to $T$.
The focusing inequality
$T\leq T_{\text{eff}}$ with $f_Q>0$ is saturated.
Therefore, in the flat FLRW background, the focusing condition
does not impose any additional constraint on the matter content. 
The result is independent of the particular form of $f(Q)$.

\section{\label{sec:Conclusion}Conclusion}
The physical acceleration of a curve should be defined using the covariant derivative associated with the connection that determines the geometry. 
In Riemannian geometry, geodesics and autoparallels coincide because the relevant connection is the Levi-Civita connection. 
Therefore, the acceleration of a geodesic vanishes with respect to the Levi-Civita covariant derivative.
In symmetric teleparallel geometry, however, the geodesic equation written in terms of the STG connection manifestly acquires a force-like contribution determined by the disformation tensor in Eq.~\eqref{eq:geodesic_in_STG}.
The force-like term would cause the departure from the force-free autoparallel curves, geodesics and autoparallels generally do not coincide.
Although the coordinate form of the linearized geodesic deviation equation 
\eqref{eq:non_covariant_geodesic_deviation_equation_in_Riemann} 
is the same as in the Riemannian case, 
its covariant form differs once the relative acceleration is defined using the STG connection, i.e., 
the relative acceleration 
$\mathrm{D}^2 n^{\rho}/\mathrm{d}\tau^2$ 
should satisfy a different deviation equation \eqref{eq:covariant_geodesic_deviation_equation_in_STG}. 
Thus, the geodesic deviation equation in STG shows physics different from the one in Riemannian geometry.
This is the subtlety that comes in the literature due to the inappropriate definition of the acceleration or other physical quantities.

The investigation of kinematics in STG is to define kinematic quantities by the deformation tensor \eqref{eq:bar-B tensor} through the nature of the velocity vector with a superscript, which differs from the literature to define the deformation tensor by using the covariant velocity vector with a subscript~\cite{Iosifidis:2018diy,Agashe:2023vsz}.
The relation between these two definitions can be found in Eq.~\eqref{eq:Bdd and Buu relation}.
For simplicity, we impose the Weyl-type ansatz \eqref{eq:Weyl-type ansatz} together with the nonmetricity-acceleration conversion equation \eqref{eq:A = 0}, 
under which we have derived the modified Raychaudhuri equation for geodesic congruence in $f(Q)$ gravity. 

In Refs.~\cite{Iosifidis:2018diy,Agashe:2023vsz}, 
the modified Raychaudhuri equation is discussed in a purely geometrical form, without introducing matter sources.
In the present work, the relation
$a^{\rho} = (1/8)Q^{\rho}$, obtained under the Weyl-type ansatz together with the nonmetricity-acceleration conversion equation
$A^{\rho} = 0$ in Eq.~\eqref{eq:A = 0},
converts the term 
${\nabla_{\!\rho}}a^{\rho}$ in 
Eq.~\eqref{eq:modified_Raychaudhuri_equation_in_STG_0} into
$(1/8){\nabla_{\!\rho}}Q^{\rho}$ in  
\eqref{eq:modified_Raychaudhuri_equation_in_STG_1}.
This makes it possible for the equation of motion to be implemented into the modified Raychaudhuri equation.
 Rewriting the equation in its trace equation 
\eqref{eq:expression_of_substitution},
we can substitute the trace equation into the last term of ${\nabla_{\!\rho}}Q^{\rho}$ to discuss the matter effect in the modified Raychaudhuri equation
\eqref{eq:modified_Raychaudhuri_equation_in_STG_f(Q)_2}.
To preserve the focusing theorem in $f(Q)$ gravity, 
we impose the focusing condition~\eqref{eq:FC_0}, equivalently
$T\leq T_\text{eff}$, where $T_\text{eff}$ is defined by Eq.~\eqref{eq:effective_trace}.
The compatibility between the Weyl-type nonmetricity
ansatz and the homogeneous and isotropic STG connection in a flat FLRW universe is analyzed. 
It requires $K_1 = K_2 = K_3$ for STG connection functions.
To satisfy the condition, only the coincident-gauge sector of Branch~1 is regular and compatible with the Weyl-type ansatz.

In the coincident-gauge sector, for the flat FLRW background, the effective trace is generically equal to the energy-momentum trace of matter, i.e., $T_{\text{eff}} = -\rho + 3p = T$ for an arbitrary function $f(Q)$ satisfying $f_Q>0$. 
Therefore, the focusing inequality $T\leq T_{\text{eff}}$ always holds as an equality and imposes no additional constraint on the matter content within this background. 
We conclude that, under the Weyl-type ansatz together with the nonmetricity-acceleration conversion equation
$A^{\rho} = 0$ 
in Eq.~\eqref{eq:A = 0}, 
the focusing theorem holds in the flat FLRW universe of $f(Q)$ gravity.

\begin{acknowledgments}
This work is supported in part by the National Natural Science Foundation of China (NSFC) under Grant No.~12547104.
\end{acknowledgments}

\appendix
\section{\label{app:Alternative derivation}Alternative Derivation}
This appendix provides an alternative derivation to obtain Eq.~\eqref{eq:The_intermediate_steps_of_Raychaudhuri_equation}.
By taking the absolute derivative of Eq.~\eqref{eq:relation from Lie_derivative} with respect to $\tau$ again,
and with the help of 
Eq.~\eqref{eq:relation from Lie_derivative} itself, 
we obtain 
\begin{align}\label{eq:second_derivative_of_deviation_vector}
    \frac{\mathrm{D}^2 \xi^\rho}{\mathrm{d}\tau^2} 
    = \xi^\alpha({\nabla_{\!\alpha}}u^\mu{\nabla_{\!\mu}}u^\rho+u^\mu{\nabla_{\!\mu}}{\nabla_{\!\alpha}}u^\rho).
\end{align}
Since STG is both curvature-free and torsion-free, the commutator of the STG covariant derivatives acting on $u^\rho$ vanishes, and, 
therefore, the Ricci identity is reduced to a symmetric relation 
\begin{align}\label{eq:the_{eff}ommutator_of_{eff}ovariant_derivatives}
    {\nabla_{\!\mu}}{\nabla_{\!\alpha}}u^\rho
    = {\nabla_{\!\alpha}}{\nabla_{\!\mu}}u^\rho.
\end{align}
By using the 
Eq.~\eqref{eq:the_{eff}ommutator_of_{eff}ovariant_derivatives}, 
the right-hand side of Eq.~\eqref{eq:second_derivative_of_deviation_vector} can be absorbed into one term, 
which can be written as
\begin{equation}
\begin{aligned}\label{eq:second_covariant_derivative_of_displacement_vector_1}
\frac{\mathrm{D}^2\xi^\rho}{\mathrm{d}\tau^2}
=\xi^\alpha{\nabla_{\!\alpha}}\frac{\mathrm{D}u^\rho}{\mathrm{d}\tau}.
\end{aligned}
\end{equation}
Furthermore, we can substitute the deformation tensor $\bar{B}^{\rho}{}_\alpha = {\nabla_{\!\alpha}} u^\rho$ in
Eq.~\eqref{eq:relation from Lie_derivative} and then take the absolute derivative with respect to $\tau$ again, which leads to 
\begin{align}\label{eq:second_covariant_derivative_of_displacement_vector_2}
   \frac{\mathrm{D}^2\xi^\rho}{\mathrm{d}\tau^2}
   = \xi^\alpha\bigg(\bar{B}^\beta{}_{\alpha}\bar{B}^\rho{}_{\beta}+\frac{\mathrm{D}\bar{B}^\rho{}_\alpha}{\mathrm{d}\tau}\bigg).
\end{align}
The left-hand side of Eq.~\eqref{eq:second_covariant_derivative_of_displacement_vector_2}
can be replaced by
Eq.~\eqref{eq:second_covariant_derivative_of_displacement_vector_1}, through the acceleration equation of the geodesic \eqref{eq:geodesic_in_STG}, 
we finally obtain
\begin{align}\label{eq:The_intermediate_steps_of_Raychaudhuri_equation_appendix}
    \frac{\mathrm{D}\bar{B}^\rho{}_\alpha}{\mathrm{d}\tau}={\nabla_{\!\alpha}} a^\rho-\bar{B}^\beta{}_\alpha \bar{B}^\rho{}_\beta.
\end{align}
Therefore, 
we show that Eq.~\eqref{eq:The_intermediate_steps_of_Raychaudhuri_equation}
can be derived directly 
without the use of the geodesic deviation equation \eqref{eq:covariant_geodesic_deviation_equation_in_STG}.

\section{\label{app:Hypersurface Orthogonality with Weyl-type Ansatz}
Hypersurface Orthogonality with Weyl-Type Ansatz}
We can easily show that, with a bit of algebra, 
the shear and rotation tensors
within the hypersurface orthogonality are reduced to the Riemannian case by imposing the Weyl-type ansatz \eqref{eq:Weyl-type ansatz}, i.e., 
$\sigma^{\mu\nu} = \mathring{\sigma}^{\mu\nu}$ 
and 
$\omega^{\mu\nu} = 0$, respectively. 

\paragraph{Shear tensor}
Eq.~\eqref{eq:sigma under ho} gives the shear tensor  
\begin{align}
    \sigma^{\mu\nu} = \mathring{\sigma}^{\mu\nu}
   - \frac{1}{2}h^{\mu\alpha}h^{\nu\beta}\,Q_{\gamma\alpha\beta}u^{\gamma}
   + \frac{1}{6}h^{\mu\nu}h^{\alpha\beta}Q_{\gamma\alpha\beta}u^{\gamma}.
\end{align}
Under the Weyl-type ansatz, the shear tensor becomes 
\begin{align}
    \sigma ^ {\mu \nu} 
    = \mathring{\sigma} ^ {\mu \nu} - \frac{1}{8} Q _ \gamma u ^ \gamma h ^ {\mu \nu} + \frac{1}{8} Q _ \gamma u ^ \gamma h ^ {\mu \nu}
    = \mathring{\sigma} ^ {\mu\nu},
\end{align}
which is the same as the Riemannian case.

\paragraph{Rotation tensor}
The rotation tensor in STG with the hypersurface orthogonality has been discussed, it is reduced to Eq.~\eqref{eq:hypersurface orthogonality for omega}:
\begin{align}
\omega^{\mu\nu} = h^{\mu\alpha}h^{\nu\beta}Q_{[\alpha\beta]\gamma}u^{\gamma}.
\end{align}
Substituting the Weyl-type ansatz, 
the rotation tensor has the form of  
\begin{align}\label{eq:rotation_tensor_under_ansatz}
\omega^{\mu\nu} &= \frac{1}{8} ( h^{\mu\alpha} h^{\nu\beta} Q _ {\alpha} g _ {\beta \gamma} u ^ {\gamma} - h^{\mu\alpha} h^{\nu\beta} Q _ {\beta} g _ {\alpha \gamma} u ^ {\gamma}) \nonumber\\[0.3em]
&= \frac{1}{8} ( h^{\mu\alpha} h^{\nu\beta} Q _ {\alpha} u _ \beta - h^{\mu\alpha} h^{\nu\beta} Q _ {\beta} u _ \alpha ),
\end{align}
in which the right-hand side is zero due to $h ^ {\alpha \beta} u _ \beta = 0$.
Thus, the rotation tensor vanishes when the Weyl-type ansatz is imposed, and we have an irrotational geodesic congruence.

\bibliography{citationlist.bib}
\end{document}